# The AgeGuess database: an open online resource on chronological and perceived ages of people aged 3-100

J. A. Barthold Jones[1*], U. W. Nash[2], J. Vieillefont[3], K. Christensen[4], D. Misevic[5¶], U. K. Steiner[1¶]

[1]Department of Biology, University of Southern Denmark, Odense, Denmark
[2]Department of Marketing and Management, University of Southern Denmark, Odense, Denmark
[3]JV Conseil Internet Consulting, Saint Nom-la-Bretèche, France
[4]Department of Public Health, University of Southern Denmark, Odense, Denmark
[5]Center for Research and Interdisciplinary, Paris, France

[*]Corresponding author
E-mail: contact@ageguess.org (JABJ)

[¶] These authors contributed equally to this work.

# Abstract

In many developed countries, human life expectancy has doubled over the last 180 years from ~40 to ~80 years. Underlying this great advance is a change in how we age, yet our understanding of this change remains limited. Here we present a unique database rich with possibilities to study the human ageing process: the AgeGuess.org database on people's perceived and chronological ages. Perceived age (i.e. how old one looks to others) correlates with biological age, a measure of a person's health condition in comparison to the average of same-aged peers. Determining biological age usually involves elaborate molecular and cellular biomarkers. Using instead perceived age as a biomarker of biological age enables us to collect large amounts of data on biological age through a citizen science project, where people upload pictures of themselves and guess the ages of other people at [www.ageguess.org](www.ageguess.org). It furthermore allows to collect data retrospectively, because people can upload photographs of themselves when they were younger or of their parents and grandparents. We can thus study the temporal variation in the gap between perceived age and chronological age to address questions such as whether we now age slower or delay ageing until older ages. The here presented perceived age data span birth cohorts from the years 1877 to 2014. Since 2012 the database has grown to now contain around 200,000 perceived age guesses. More than 4000 citizen scientists from over 120 countries of origin have uploaded ~5000 facial photographs. We detail how the data are collected, where the data can be downloaded free of charge, and the contained variables. Beyond ageing research, the data present a wealth of possibilities to study how humans guess ages and to use this knowledge for instance in advancing and testing emerging applications of artificial intelligence and deep learning algorithms.

# Introduction

Global record life expectancy—the highest average life span among the world's populations—has risen steadily by 3 months per year for the last 180 years [1]. Underlying this exceptional increase is a change in how we age, even though our understanding of this change remains limited. Reaching this understanding is not only a pressing concern for anticipating future changes in life expectancy, a question of high social and economic importance, but also lies at the core of our human self-understanding. Our lives will end with death and most of us have an interest in knowing when this time might come. Here we present a unique open-access database, rich with possibilities for studying ageing: the AgeGuess.org database on people's perceived ages (i.e. how old someone looks to others) and chronological ages. The data on perceived age, an established biomarker of biological age [2], originate from a citizen science project, where people upload pictures of themselves and estimate the ages of other users.

With rising life expectancies and commonly associated plummeting fertility rates, older people will steadily increase as a proportion of the population—a phenomenon known as population ageing [3]. By 2050, 1 in 10 citizens of the European Union will be aged 80 years or older — a towering tenfold increase in 10 decades [4]. Population ageing heralds a suite of challenges for the economy, social security, and health care of countries. Therefore, organisations like the United

Nations and World Health Organization work to increase awareness of how population ageing will affect people's lives and the lives of their children and grandchildren [5]. Finding solutions to the challenges of ageing populations is a pressing task also recognized by national and international research and industry funding agencies [6,7]. Understanding how the human ageing process has changed over the last 100 years, and to predict how it will continue to change in future, is an important piece of the puzzle.

The global burden of age-related disease will rise exponentially if we live longer but spend our older years with loss of functioning and dignity [8]. Therefore, to anticipate future health care needs, one focus of ageing research is whether we are not only living longer, but also better. Or do we simply spend more years in a degenerative state marked by ill health and disability [9]? Research results overall points towards a positive answer: limitations and disabilities of people aged 85 or younger seem to indeed be postponed to create longer lives in good health. However, conclusions remain tentative not least due to 1) inconsistency among studies in disability markers, survey questions, and participation rates, 2) an overall increase in early diagnosis of chronic diseases and conditions, and 3) a general exclusion of the institutional population including care home inhabitants. Incidence and prevalence studies of old-age disease and disability have a further shortcoming with respect to revealing changes in the human ageing process: these studies concentrate on the oldest-old, a category of people who are typically already affected by age-related diseases.

Therefore, to understand how ageing is changing, research must widen its focus from the oldest-old to studying ageing across all ages. There is good evidence to support this argument: age-related changes to physiology, building up to age-related diseases, accumulate from early life [10,11]. Young adults already differ in their biological ages [12]. Biological age describes the relative condition of, for example, cardiovascular, metabolic or immune system, biological ageing is therefore a change in functioning over time. It is usually determined by measuring an array of biomarkers of molecular and cellular events, which then are compared to a cohort average to determine biological age [13].

Obtaining the necessary data on biological age is challenging. Biomarkers of ageing are as complex as the biological phenomenon itself [14,15]. From recent research, perceived age emerges as an excellent candidate biomarker for biological age for studies that require large amounts of data [2,12,16,17]. This kind of data can reveal how the ageing process changes among birth cohorts over time, and its manifestation within and between individuals over time.

Here we present a wealth of perceived age data spanning birth cohorts from the years 1877 to 2014. The data were collected by a citizen science project since 2012, where citizens upload facial pictures of individuals with known age and guess ages of other users. The data collection is ongoing. More than 4000 citizen scientists from over 120 countries of origin have uploaded ~5,000 pictures (Fig 1, Fig 2a). The citizen science project continues to grow steadily (Fig 2b). Communication by the media and outreach activities on social media (Facebook: www.facebook.com/ageguess.org/, Twitter: @ageguess.org) and in person (e.g. open science days) aim to both recruit more users and to inform the public about the change in how we age.

In the following, we introduce how we collect the data via a webpage, and how we recruit further citizen scientists to the project. We then describe the database in detail, both providing an extensive summary of the data and information on the database variables. Finally, we suggest areas of research, which could exploit the database and make it available for download, free of charge.

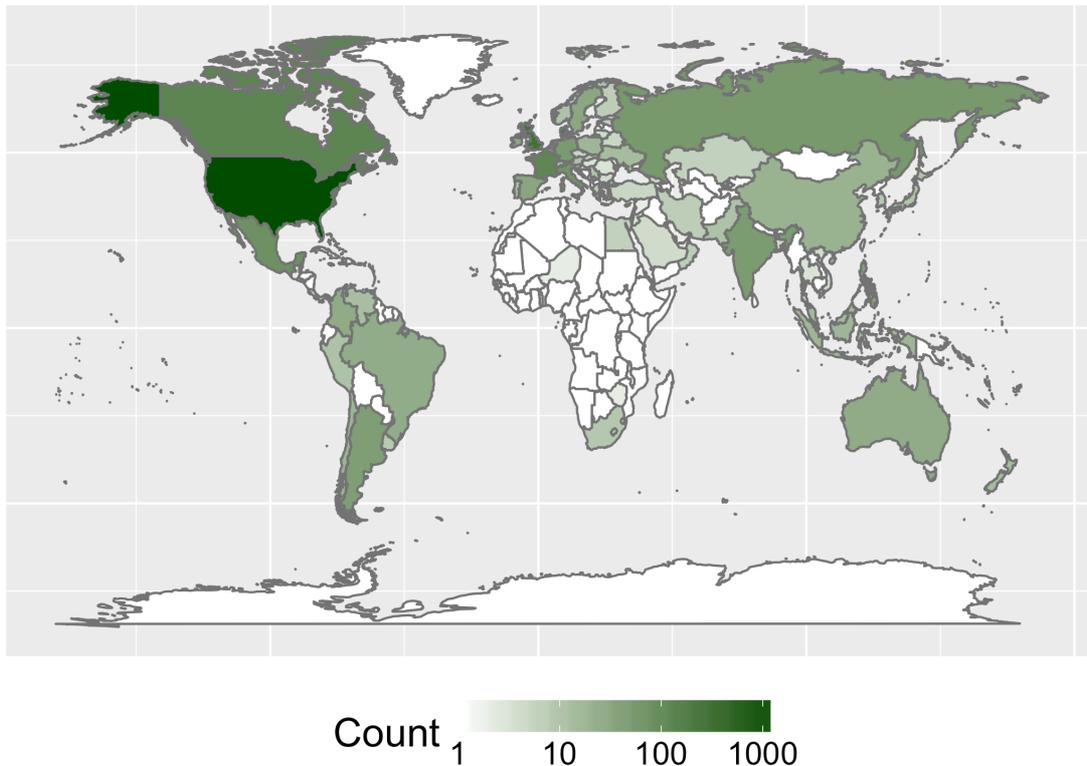

**Fig 1. AgeGuess users by country of origin**

# The webpage

## Organisation

The webpage at www.ageguess.org is the platform we use to collect the perceived age data. The webpage is licensed under a Creative Commons Attribution-NonCommercial-NoDerivatives4.0 International License. The webpage is hosted in France and has been approved by the Commission nationale de l'informatique et des libertés (CNIL, National Commission on Informatics and Liberty; declaration #1800944v0), France's regulatory body ensuring the application of data privacy law to the collection, storage, and use of personal data. The webpage is built with Drupal (www.drupal.org), a free and open-source content management system (CMS) based on PHP and MySQL. Drupal provides login/logout functionalities and account activation and deletion. While users retain the copyright of their pictures, the AgeGuess project owns the rights to use content on the webpage covered by intellectual property rights, including text, images, graphics, logos, icons, sounds, and software. The full terms and conditions are available at www.ageguess.org/legal-notice. Below we describe the most prominent features of the webpage regarding data collection. Further details about both the front and back-end behaviour of the webpage can be received upon request.

# Data collection

To contribute to the data collection, users create an account with a verified email address, link to or upload photographs of themselves, and provide basic information about themselves: birth year, age in the photograph, ethnicity, and birth country. They can then proceed to guess the ages of other users in these users' photographs. After each guess, the users see the real age of the person in the picture, summary statistics of previous guesses of other users, and a histogram of previous age guesses. Similarly, the user can see this information for their own photographs on the user's personal account page. As an additional feature, users can also upload photographs of other people (e.g. relatives) for which the user owns the copyright or that are available under Creative Commons license. The users provide the same basic information for the persons in the photographs as they provided for their own photographs and specify their relationship (i.e. friend or family member). The minimum legal age to open an account is 14 years.

To provide a further incentive to contribute, the webpage is set up as a simple online game. For uploading pictures and for each guess, the users receive a number of points depending on the accuracy of the guess (exact guess 10 points, 1-2 years off 7 points, 3-5 years off 5 points, 6-10 years off 2 points, more than 10 years off 1 point, 10 points for each uploaded or linked picture). On their personal account page, the user can see the cumulative number of gained points, the number of guesses made, and the proportion of fully accurate guesses. At [www.ageguess.org/ranking](www.ageguess.org/ranking) users can compare how good they are at guessing ages in comparison to other users by point scores and by the mean deviance of their age guesses from the real age of displayed persons. Users can locate their own position on ranking lists by clicking a provided button.

The webpage displays photographs to users for guessing following a specific algorithm, which creates a queue, an ordered list of photos that will be shown to users to guess. To be eligible to be picked by the algorithm for display to a user, pictures have to fulfil the following criteria: 1) not be uploaded by the user themselves to prevent users rating their own images, 2) not being guessed and/or seen already by the user, and 3) not being reported more than four times. Users can report pictures when making guesses by clicking one of the options: rotation needed, cropping needed, missing person, more than one person, copyright infringement, and offensive content such as nudity or violence. Taking into account these criteria, the algorithm sorts the pictures ascendingly by number of previous guesses, placing photos with fewer guesses at a top of the list. Pictures with the same number of guesses are chosen at random. Furthermore, the user can skip a photo via a "skip" button, for example to avoid guessing ages of people the user knows. The user can however not skip more than 4 times per session. The system renews the queue each time a user logs in.

It is worth noting that features of the algorithm and the overall system were gradually refined over time with growing knowledge on users' behaviour and potential problems. A suite of countermeasures against malicious users are in place. For example, to reduce the number of malicious users, who may upload unauthorised pictures, try to get access to pictures of other users, and/or troll the webpage operations, the webpage requests email validation at registration. We conceal full details of these measures in order to keep malicious users at a disadvantage.

# Obtaining the AgeGuess data

Users with an active AgeGuess account can download the data described in the next section from a repository at www.ageguess.org/download. The AgeGuess project updates the publicly available data every three months. We urge users to read the user agreement and to cite this paper when using the data. We do not publish the photographs open access to respect EU privacy protection law. All data are fully anonymised and any attempt to reveal the identity of users violates AgeGuess regulations.

# The database

## History and internal organisation

In 2012 U. Steiner and D. Misevic started the AgeGuess.org citizen science project. They form the core committee and are responsible for creating and updating protocols for data collection and for the overall infrastructure of the database, as well as for securing funding. They are supported by webpage building and database construction expert, J. Vieillefont, who created and maintains the current version of the webpage and the database. The first fully functional version of AgeGuess.org was coded by Charlotte Le Pesquer. Furthermore, a team of scientific advisors stretching both academic disciplines (e.g. public health) and the industry (e.g. pension providers) helps shape the scientific directions of the project and highlights funding opportunities. Depending on availability of funding, one or more pre- or post-doctoral fellows have worked on data analysis and outreach.

## Variables and descriptions

The data are stored in a MySQL database but are accessible to the public as five csv files. They respectively contain information on *Guess*, *Photos*, *Gamers*, *Quality*, and *Report*, using those names and .cvs extensions. In the following, we describe the variables in each of the csv files. All missing data are encoded as NA.

The *Guess.csv* file stores the information regarding the photographs using the following variables: *guess_id*, *photo_id*, *uid*, *ageG*, *outG*, *access*, *access_datetime*, *repeated_guess*. The *guess_id*, *photo_id*, *uid* variables contain the individual identifiers of each guess, the photograph guessed on, and the user who made the guess. The *ageG* and *outG* variables describe the guessed age and the deviation in the guess from the real age in years, respectively. The access and access_datetime variables store the timestamp when the guess was made in Unix time and in date and time UTC+1:00, respectively. Finally, the *repeated_guess* variable is an indicator variable (TRUE/FALSE) for separating unique from repeated guesses from the same user on the same photograph. Some research questions might use this information, for others the data can be subsetted using *repeated_guess*.

The *Photos.csv* file stores the information regarding the photographs using the following variables: *photo_id*, *uid*, *age*, *relation*, *gender*, *ethnicity*, *birth_country*, *birth_year*, *death_age*, *created*, and *created_datetime*. The *photo_id* and *uid* variables represent the individual identifiers of the photograph and the user who uploaded the photograph. The *relation* variable indicates whether the photograph is of the user or of another person to which the user has a relation (current categories: user, grandchild, half sibling, maternal grandparent, mother/father, paternal cousin, paternal grandparent, paternal uncle/aunt, sibling, son/daughter, unrelated or friend). The *gender*, *ethnicity*, *birth_country*, *birth_year*, *death_age* variables contain the respective basic demographic information for the person in the photograph. The *created* and *created_datetime* variables store the timestamp when the photograph was added in Unix time and in date and time UTC+1:00, respectively.

The *Gamers.csv* file stores the information regarding the users (aka gamers) with the following variables: *uid*, *g*, *ng*, *points*, *gender*, *ethnicity*, *birth_country*, *birth_year*, *access*, *created*, *access_datetime*. These variables store the individual identifier of the user (*uid*), the number of correct guesses the user made (*g*), the number of other guesses (*ng*), and the points gained in the online game (*points*). Furthermore, the file contains the users' basic demographic information regarding gender, ethnicity, birth country, and birth year, stored in variables of these names. Finally, the *access* and *access_datetime* variables store the timestamp when the user last logged in Unix time and in date and time UTC+1:00, respectively, while the *created* variable represents the timestamp of when the user made an account with AgeGuess.

The *Quality.csv* file contains information on quality reports that users have made on photographs. The variables are *quality_id*, *photo_id*, *quality*, *uid*, *created*, *created_datetime*. The *quality_id*, *photo_id*, and *uid* variables contain the individual identifier of the quality assessment, the photo on which the assessment was made, and the identifier of the user who made the assessment, respectively. Quality itself is encoded as 1 = high, 2 = medium, 3 = low in the *quality* variable. The timestamps of the assessment in formats described above are stored in the *created* and *created_datetime* variables.

Finally, the *Report.csv* file pertains to information on any other reports made on photographs. The variables are *report_id*, *photo_id*, *uid*, *comment*, *created*, and *created_datetime*. The *report_id*, *photo_id*, and *uid* variables store the individual identifiers of the report, the photograph on which the report was made, and the identifier of the user who made the report, respectively. Report categories are *rotation needed*, *cropping needed*, *none or more than one person*, *copyright infringement*, *offensive content*, and combinations thereof. The AgeGuess team regularly edits images after report, for example when cropping is needed, and retains the edited photographs if suitable. Photographs and data associated to the other reports categories are deleted. Finally, after internal checks the system adds reports related to missing photographs and inaccurate data on birth year and age. The timestamps of the report in formats described above are stored in the *created* and *created_datetime* variables.

## Data quality and outlier removal

The data set can contain both false and missing data that may have been entered by users either by mistake or intentionally. Therefore, we perform some basic data cleaning steps before publication of the data. From the *Guess* data we delete all guesses that are more than two times the standard deviation away from the mean age guess on a photograph. We further remove all guesses on photographs with less than 10 guesses on the photograph. Using the information in the *Report* data (see paragraph above), we delete guesses on photos with inaccurate age or birth year. Since not all inaccurate birth years are flagged by internal system checks, we replace in both the *Photos* and *Gamers* data all unrealistic birth years (< 1800 or >2018) with NA. The whole, uncleaned data set can be obtained upon request.

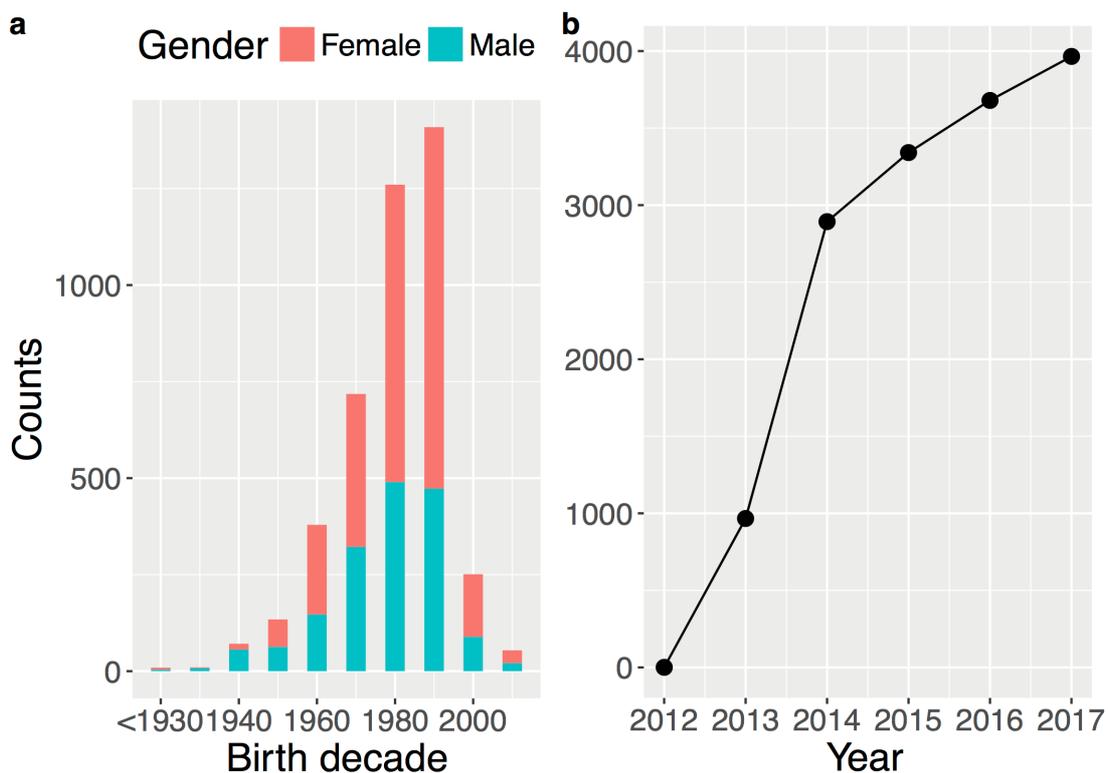

**Fig 2. Summary of perceived age data in the AgeGuess.org database. a** Number of photographs by sex and birth decade of individual. **b** Cumulative numbers of users.

Furthermore, the data quality of the AgeGuess.org database experiences a trade-off common to citizen science projects, where large quantities of data are won at the expense of representativeness of sample and data accuracy. Neither the AgeGuess users nor the persons displayed in the photograph are representative samples of the population with respect to age, geographic location, or ethnicity. Furthermore, the uploaded photographs are not standardised with respect to posture, lighting, face expression, clothing, background, distance to camera, hairstyle or dye, make-up, or the use of accessories such as hats, jewellery, or glasses. Some of these factors may be used to deliberately conceal age: older adults may use particularly make-up and hair dye to appear younger, while younger adults may manipulate their looks to appear older. Finally, we discourage editing photographs to alter the age appearance and offer a report option to flag those photographs. However, some manipulated photographs may have undergone unnoticed.

## Data summary

After running the data cleaning protocol, AgeGuess has, as of late February 2018, 4010 users from ~120 countries of origin of which 2099 are female, 1573 male, and the rest is unknown (Fig 1). These users have uploaded 4335 photos of 2634 females and 1701 males (Fig 2). The age of the persons displayed in the photographs ranges from 3 to 100 years old. The earliest and latest corresponding birth years were 1877 and 2014, respectively. The data contain repeated measures on 467 individuals with more than 215 individuals having uploaded three or more pictures of themselves.

Overall users have guessed ages 180,798 times. We have at least 10 repeated guesses for each photograph, with a maximum of 532 repeated guesses and a median of 34 guesses. The variation in number of guesses stems from earlier versions of the photograph-selecting algorithm, which did not take into account the number of previous guesses on a photograph. The deviation of the mean age guess from real age for each photograph is normally distributed with a mean close to 0 (Fig 3).

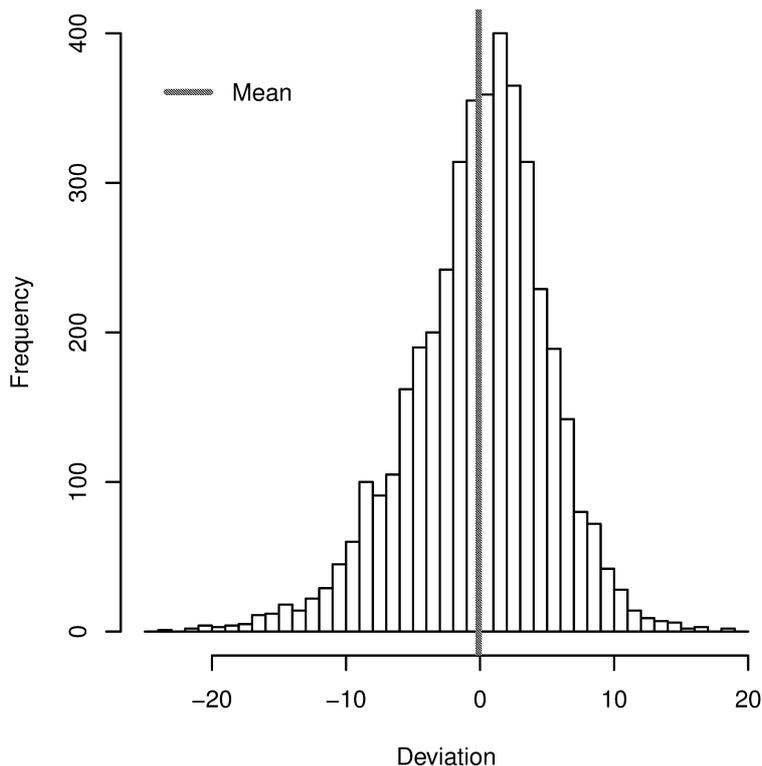

**Fig 3. Histogram of the deviation of the mean age guess from real age for each photograph**

# Research opportunities

The AgeGuess data provide exceptional opportunities to approach research questions across scientific fields. The research opportunities broadly split into two main directions. The first one relates to ageing research and some of the questions outlined in the introduction, e.g. evaluating ageing processes by studying temporal variation in biological age and the difference between perceived and chronological age. The second direction relates to a more sociological view, where the users themselves are the subject of study, and research questions center around their ability to guess ages.

Regarding the first research direction, the collected data on perceived age and chronological age can inform on the basic ageing process. The data can reveal whether more recent birth cohorts are biologically younger than their earlier counterparts, e.g. is a 40-year-old today biologically younger than a 40-year-old in the 1980s, and does that difference hold in the same way for 30-year-olds? We can investigate how shifts in biological age over time accompany shifts in life expectancy [12,18]. Research questions can also focus on how ageing happens within a lifetime, i.e. do we age continuously throughout our lives or are there boosts and arrests of ageing? Furthermore, quantitative geneticists can exploit relatedness among persons in uploaded photographs to study diverse questions, including to what degree biological age is heritable, and how heritability of biological age relates to known heritability of lifespan. Computer scientists can use the data to train and evaluate age estimation algorithms [19]. Access to the original images requires a close collaboration with the AgeGuess project due to new EU regulations on privacy of identifiable pictures of persons.

The second research direction centers around the users and their ability to guess ages. General questions about the ability to guess the age of people include: 1) are we better at guessing the age of our own age group, 2) does our ability to guess ages increase with experience (age or number of previous guesses on AgeGuess), and 3) are we better at guessing the age of our own sex or the opposite sex, and is the age of one sex easier to guess than the other? Such questions can test hypotheses derived from sexual selection theory, for example in the context of human partner choice. Furthermore, studying whether it is easier to guess the age of people of our own ethnicity compared to other ethnicities might provide information on generalities of ageing processes and commonalities of ageing signs. The data might furthermore reveal guessing abilities related to exposure to specific ethnic or age groups. For example, opposite to expectation geriatric nurses, who daily work with elderly patients, are not better at guessing ages of elderly citizens than male undergraduate students [2].

As focus turns from the subject in the photograph to the judgment, the AgeGuess data become highly valuable for answering basic questions about information processing. More specifically, the data provide an outstanding opportunity to investigate to what extent hidden aspects of the environment can be measured using perceptible cues that correlate more or less with these unobservable parts. This problem of inferring hidden aspects is, and has been, fundamental to human survival, and is therefore important for cognitive psychologists and neuroscientists alike [20]. Moreover, in their development of artificial intelligence, the cue learning problem is, with some controversy, being studied by computer scientists [21]. How much information about a person can be extracted from the person's facial features, and to what extent it matters how different features of the face are combined in making the judgment [22] are interesting questions. They are also questions with potential to challenge the organisation of society, in the event judgments are done at mass scale by computers with live access to extensive networks of cameras [23].

The questions that we list here are certainly not exhaustive and are meant only as an illustration of the power of the data for answering broad scientific questions. We believe many more questions can be addressed using the AgeGuess data, including those that we as data collection initiators have not thought about and may not be able to imagine. We invite researchers and citizens alike to tap into this rich resource and provide open access to the data free of charge.